\documentclass[a4paper]{jpconf}
\usepackage{graphicx}
\usepackage{bm}
\begin{document}
\title{Support Vector Machines and generalisation in HEP}

\author{Adrian Bevan, Rodrigo Gamboa Go\~ni, Jon Hays, Tom Stevenson}

\address{School of Physics and Astronomy, Queen Mary University of London, Mile End Road, London, UK}

\ead{a.j.bevan@qmul.ac.uk}

\begin{abstract}
We review the concept of Support Vector Machines (SVMs) and discuss examples of their
use in a number of scenarios.  Several SVM implementations have been used in HEP and we
exemplify this algorithm using the Toolkit for Multivariate Analysis (TMVA) implementation.
We discuss examples relevant to HEP including background suppression for $H\to\tau^+\tau^-$ at the LHC
with several different kernel functions.  Performance benchmarking leads to the issue of 
generalisation of hyper-parameter selection. The avoidance of fine tuning (over 
training or over fitting) in MVA hyper-parameter optimisation, 
i.e. the ability to ensure generalised performance of an MVA that is independent of the 
training, validation and test samples, is of utmost importance.  We discuss this issue and 
compare and contrast performance of hold-out and k-fold cross-validation.  We have extended
the SVM functionality and introduced tools to facilitate cross validation in TMVA and
present results based on these improvements.
\end{abstract}

\section{Introduction}
\label{sec:other}

These proceedings discuss High Energy Physics (HEP) usage of Support Vector Machines (SVMs)~\cite{svm}, and in particular 
recent improvements to the functionality of the TMVA~\cite{TMVA} implementation.  These improvements are deployed in the
ROOT git repository and from version 6.08. Having discussed this 
Machine Learning (ML) algorithm we proceed to raise the issue of generalisation of hyper-parameters (HPs); and in particular 
the use of hold-out and cross-validation (CV).  We also discuss the issues of understanding variance on the classification 
and performing a hypothesis test to address the issue of identifying if the HPs of a Multivariate Algorithm (MVA) are generalised,
having used some method to promote HP generalisation.

\section{SVMs}
\label{sec:svm}

\begin{figure}[!th]
\begin{center}
\includegraphics[width=0.45\columnwidth]{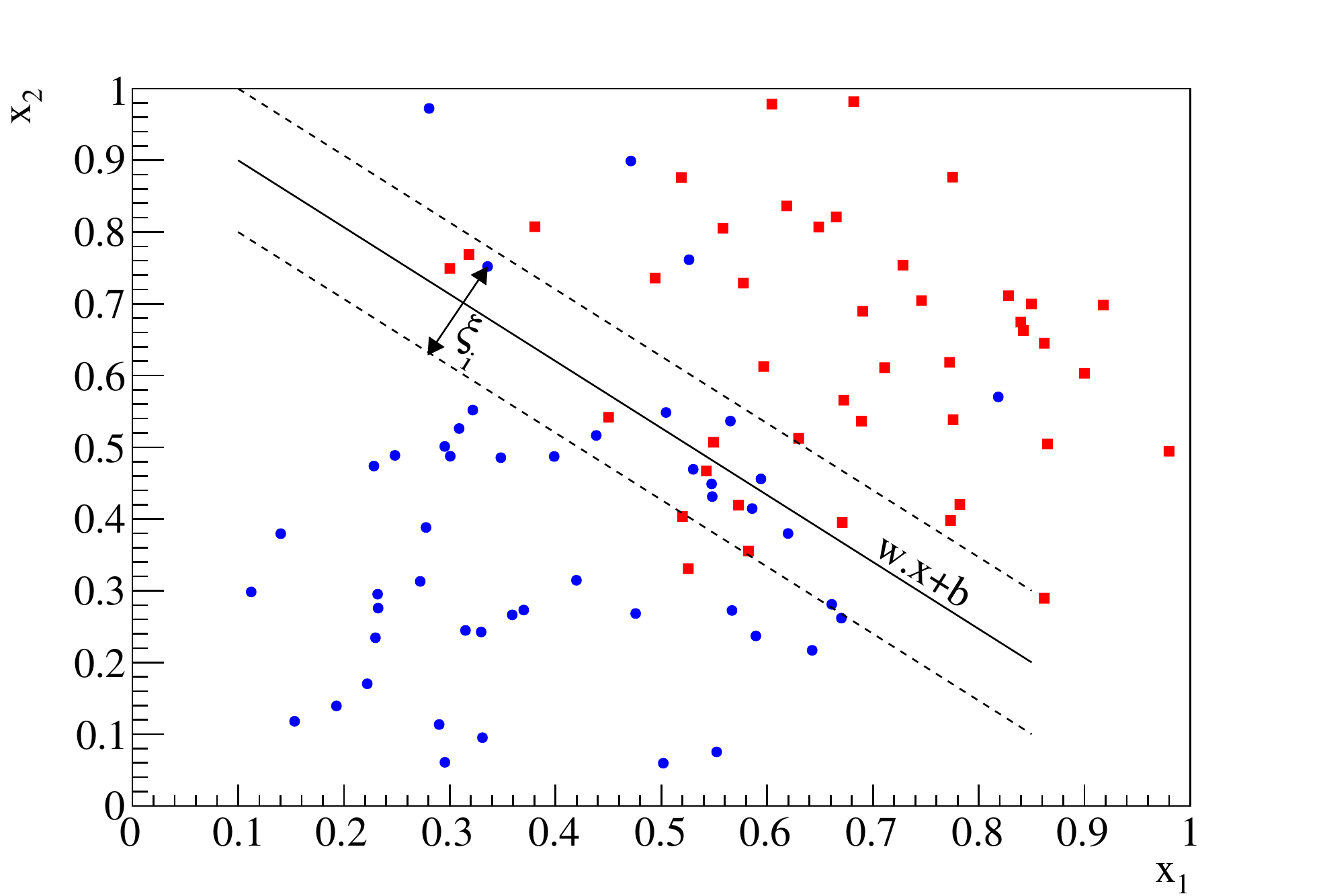}\hspace{2pc}
\begin{minipage}[b]{0.95\columnwidth}\caption{\label{fig:softmargin}Soft margin SVM showing the (solid) maximal margin hyper-plane
and (dashed) functional margins for two categories of events (dots and squares) in two dimensions.}
\end{minipage}
\end{center}
\end{figure}

SVMs were developed in the 1960's in a hard margin form.  These were based on Rosenblatt's perceptron that
allows one to trivially separate distinct types of data using a linear decision boundary, given
by ${\bm w} \cdot {\bm x} + b$~\cite{Rosenblatt1958}.  Here ${\bm w}$ is a set of weights defining the orientation of the 
separating hyper-plane, $b$ is the bias (offset from the origin) and ${\bm x}$ is an event.
Subsequent developments expanded the power of SVMs in two significant ways; firstly to allow for misclassification of 
events and thereby extending the algorithm to a broader set of problems, including those encountered in a typical HEP analysis. 
This extension requires introduction of an overall cost parameter $C$ for event mis-classification as well as
slack parameters $\xi_i$ for the mis-classification of any given event (or support vector, SV). 
Here $C$ is a HP that alters
the selection of the hyper-plane separating classes of event. This type of SVM is 
referred to as a Soft Margin SVM (see Fig~\ref{fig:softmargin}). The second extension was to replace 
the use of scalar products between two SVs, 
$K({\bm{x}_i}, {\bm{x}_j}  ) = \bm{x}_i \cdot \bm{x}_j$, with inner products 
in terms of kernel functions (KFs) of SVs, 
$K({\bm{x}_i}, {\bm{x}_j}  ) = \langle \phi({\bm{x}_i}) , \phi(\bm{x}_j)\rangle$.  These 
functions introduce additional HPs (KF parameters\footnote{Sometimes $C$ is also referred to as a KF parameter.}) 
that need to be determined 
in order for one to separate classes of event.  The use of KFs allows us to attempt to linearly separate types of 
event in a dual hyperspace of features $F$ mapped implicitly from the input feature space $X$. It is not
necessary to understand the mapping $X \mapsto F$.  For this mapping to exist a KF needs to be symmetric 
and satisfy Mercer's condition~\cite{mercer}.  For such kernels the algorithm can be used without detailed 
knowledge of the transformation; this is referred to as the \emph{kernel trick} in the literature.
While the 
kernel trick allows us to use complicated KFs it does obscure understanding of  
how the separating hyper-plane is determined.  
This plane is determined for a 1-norm SVM by optimising the dual space Lagrangian
\begin{equation}
\widetilde{L}(\alpha)=\sum_{i=1}^{n}\alpha_{i} - \frac{1}{2}\sum_{i,j}^{n}\alpha_{i}\alpha_{j}y_{i}y_{j}\left\langle\mathbf{x}_{i}\cdot\mathbf{x}_{j}\right\rangle,
\label{eq:hardMarginLagrangian}
\end{equation}
to determine the Lagrange multipliers, $\alpha_i \in [0, C]$, in order to maximise the margin.  
The Lagrange multipliers are
determined using Sequential Minimal Optimisation (SMO) and the constraint equation $\sum_{i=1}^{n}\alpha_{i}y_{i} = 0$.
Here the $y_i$ are example labels given by $\mathrm{sign}( \langle {\bm w} \cdot {\bm x}_i\rangle + b)$.  
They denote which side of the decision boundary an event lies on.  The functional margin of an example $({\bm x}_i,y_i)$ 
is given by $\gamma_i = y_i(\langle {\bm w} \cdot {\bm x}_i\rangle + b)$ where positive values are obtained for correctly
classified examples.
The $\alpha_i$ and SVs determine hyper-plane parameters as ${\bm w}$ and $b$ are given by
\begin{eqnarray}
{\bm w} &=& \sum\limits_{i=1}^n \alpha_i y_i {\bm x}_i,\\
b &=& -\frac{\max_{y_i=-1}(\langle {\bm w}\cdot {\bm x}_i\rangle) + \min_{y_i=1}(\langle {\bm w}\cdot {\bm x}_i\rangle)  }{2}.
\end{eqnarray}

SVMs are widely used outside of physics; for example in handwriting recognition, speech recognition, bioinformatics,
risk analysis, finance, electronics, etc.  Many of these problems involve distinguishing 
between features that are largely not overlapping; for example the classification of handwriting
as illustrated by the MNIST classification example in TensorFlow~\cite{LeCun1998a}.
Practitioners typically apply different KFs to a given problem in order to empirically validate that a given 
KF provides good performance for that problem.  In doing so one implicitly relies on the kernel trick to work without
trying to understand details of the problem in the dual feature space.
The most commonly used KFs in HEP examples are scalar products and the radial basis function (RBF)
where performance has been reported to be comparable with that for other ML
algorithms; Neural Networks (NNs) and Boosted Decision Trees (BDTs).
In contrast to other fields there has not been a broad uptake of SVMs within the HEP community. 

Several reports of SVMs use in HEP have been made: including
jet flavour tagging and muon identification~\cite{Vannerem:1999wm},
top physics~\cite{Vaiciulis:2002jw,Whitehouse:2010zz}, background suppression for $W\to e\nu$
signatures~\cite{Sforza:2013hua}, and SUSY searches~\cite{Sahin:2016qgg} at
LEP, Tevatron and LHC experiments.  Ref.~\cite{Whitehouse:2010zz} discusses a custom
implementation of a SVM, and where specified the other references rely on
the libsvm implementation~\cite{CC01a}.
A review of SVM use in HEP as of 2008 can be found in Ref.~\cite{Vossen:2008se}.
We have applied SVMs to Higgs physics~\cite{kaggle}, 
using the Kaggle Higgs Boson ML Challenge data sample~\cite{ACATprocs}. 
The examples discussed below use the 
TMVA SVM algorithm implementation that uses SMO for Lagrange multiplier 
optimisation and Minuit~\cite{minuit} for hyper-parameter optimisation.

\subsection{Kernel Functions}
\label{sec:svm:kernel}

KFs used in previous HEP examples are scalar products and the RBF function. The latter is
\begin{eqnarray}
K({\bm{x}_i}, {\bm{y}_i}  ) = e^{-\Gamma ||{\bm x}_i - \bm{y}_i ||^2},
\end{eqnarray}
where $\Gamma = 1/2\sigma^2$ is a HP to be optimised.  Having just a single HP to determine (in addition
to $C$) has the advantage of being a relatively cheap computing task. The disadvantage of the RBF function is
that it assumes that the norm of the SV is sufficient to distinguish between events of different categories.  
Thus information may be lost in the computation of the KF for some problems.  
Computer Science literature discusses a number of other KFs, some of which
we have added to TMVA. These include multi-Gaussian (MG) and polynomial (Poly) functions
\begin{eqnarray}
K({\bm{x}_i}, {\bm{y}_i}  ) = \prod\limits_{j=1}^{\dim ({\bm x}) } e^{-\Gamma_j (x_{ij} - y_{ij})^2 }, \,\,\,\,
K({\bm{x}_i}, {\bm{y}_i}  ) = \left( \sum\limits_{j=1}^{\dim ({\bm x}) } x_{ij} y_{ij} + c\right)^d,
\end{eqnarray}
respectively.  The MG KF has $\dim({\bm x})$ HPs, $\Gamma_j$, to overcome potential loss of 
information that the RBF may result in.  The HPs of the Poly KF are $c$ and $d$.
The sum and product of KFs are also valid KFs and these four functions are now available in TMVA.

\subsection{Hyper-parameter optimisation}
\label{sec:svm:parameters}

Logarithmic grid searches have been used (via libsvm and SVM-HINT) in order to determine HPs 
for RBF KFs.  This is computationally inexpensive way to determine
HPs for a KF with a low dimensionality, especially if it is of the form of an RBF or MG kernel.  
The error in selecting optimal HPs is governed by the 
granularity of the lattice used in the grid search.
As soon as one moves to a high dimensionality of 
KF HPs this optimisation paradigm becomes expensive as the grid search scales with 
$\dim({\bm x})$.  To address this issue for TMVA the Davidon-Fletcher-Powell based Minuit optimisation 
algorithm~\cite{minuit} has been adopted for HP optimisation (HPO)~\footnote{A grid search option is also 
being implemented for future release.}.  HPO using this approach is more computationally
expensive than the grid search approach for the 2D RBF optimisation problem however in general this
provides a more accurate solution and is not specifically attuned to RBF and MG kernel functions, unlike the 
logarithmic grid search.

\subsection{Other remarks}
\label{sec:svm:other}

Anecdotal remarks related to the use of SVMs note that these tend to result in a more generalised solution
than other methods for a given sample size of training data.  This is plausible and can be seen trivially
for a hard margin SVM following the observation that the maximal margin hyper-plane is defined only by the 
SVs; hence only those, generally few, events on the decision boundary edges define the hyper-plane.
It is not trivial to translate this observation to a soft margin SVM, however in general a parallel
situation will arise, where now it is the set of SVs with non-zero $\xi_i$ that determine the 
definition of the decision boundary.  The SVs with large values of $\xi_i$ dominate the 
determination this boundary.  In practice we observe better generalisation performance for 
SVMs with small training example sets compared with decision trees and neural networks in the
comparisons that we have performed.
While it is possible to optimise the HPs for a SVM, the issue of parameter generalisation remains 
one to be addressed as this is not guaranteed.  Generalisation bounds exist for SVMs and methods 
used to promote generalised solutions are discussed further in Section~\ref{sec:generalisation}.

As mentioned above the hard margin SVM requires determination of the handful of SVs that fall on
the functional margin to determine the HPs and hence decision boundary.  Extending this to 
a soft margin SVM problem like those encountered with HEP background fighting problem the number of SVs 
routinely increases to encompass all training data.  This means that the $N_{SV}\times N_{SV}$ Gramm matrix 
to be computed scales from a low dimensionality to a high dimensionality and becomes computationally
expensive relative to training many other algorithms.  
This can be seen trivially by the computing time required to fit a SVM, vs that to 
train a NN or BDT using TMVA for a given data sample.  
Many of the SVs for a HEP soft margin problem have slack parameters that are small and therefore
they do not significantly contribute to the determination of the maximal margin hyper-plane.  The low
$\xi_i$ SVs are largely superfluous to the problem and a similar performance is attainable by presenting
the SVM with a redacted data set that focuses on the SVs with $\xi_i$.  In itself this observation is
of little use as one needs to perform the SMO optimisation of the Lagrangian in order to determine the 
$\xi_i$'s and in doing so therefore fix the Gramm matrix dimensionality accordingly.  Data redaction
or sparsification can be achieved by noting that the SVs with low value $\xi_i$'s are far from the decision 
boundary.  Hence if a-priori it is possible to identify regions far from the decision boundary then one
can decrease compute cost by removing those regions from training at the expense of introducing an
approximation to the definition of the maximal margin hyper-plane.  

\section{Generalisation}
\label{sec:generalisation}

Obtaining a set of hyper-parameters for an MVA that can provide predictive performance when applied to 
any unseen data sample consistent with the hyper-parameter optimisation is desirable.  In HEP it is typical
that generalisation is promoted using the hold-out technique (see below).  
HEP problems include both regression and classification variants.  For classification we need to compare
classification rates (or confusion matrices) 
and for regression problems we need to compare histograms to test for over-training.
In practice a $\chi^2$ by eye or Kolomogrov-Smirnov (KS) test is performed on training and test histograms
for the MVA output.  In the case of TMVA the latter uses a binned implementation in the TH1 class that
is known to be incorrect~\cite{th1}.  Furthermore the correctly calculated 
(un-binned) KS test statistic compares the similarity of
the bulk of a distribution which is not the right comparison to make in general.  This can be 
seen clearly for examples of searches for new particles in tails of distributions that corresponds
to a large proportion of results in HEP at this time.  
The issue of similarity metrics is discussed elsewhere~\cite{similarity}
and it is evident from that review that no single metric provides a general solution to this issue.

We now consider three issues: firstly enhancing generalised behaviour at the 
HPO stage of algorithm tuning; secondly to determining if the set of HPs corresponds 
to generalised performance; thirdly to determine which MVA provides the best performance for 
a problem.  Regarding the first point we focus on the hold-out (Section~\ref{sec:generalisation:holdout}) 
and k-fold CV (Section~\ref{sec:generalisation:cv}) 
methods that are generally applicable to MVAs. 
Note that NNs use regularisation~\cite{regularisation} and deep networks utilise dropout~\cite{dropout} and
maxout~\cite{maxout} techniques.  The second issue revolves around estimation of bias and variance
for the ML algorithm and performing a hypothesis test to reject the null hypothesis that
the HPs are generalised (Section~\ref{sec:generalisation:other}) in order to choose an MVA model.
The third issue can be dealt with in a number of ways and in Ref.~\cite{ichep2016} 
we use a likelihood ratio test of a proxy of the observable being measured in order to illustrate
this issue.

\subsection{Hold-out method}
\label{sec:generalisation:holdout}

The ``standard'' method currently in use within HEP for promoting generalised HP determination for supervised learning 
algorithms is the hold-out CV method~\cite{Devroye1979}.  This involves dividing control (Monte Carlo simulated or real) 
data samples into two; taking one sample
used for HPO (the training sample) and reserving the second sample to evaluate the performance
of the HPs (the test sample).  Some similarity test of the classification performance or binned distributions of
the algorithm score can then be used to gauge if the HPs are generalised; however it is evident that this crucial step
is sometimes overlooked.  When this last step is not overlooked then often an incorrect binned KS test is performed
as noted above~\footnote{To rely on the KS probability 
reported by the TH1 implementation to accept/reject a given training will lead to rejection of perfectly good trainings,
which is to be avoided.}.
Having performed the hold-out validation of HPs and assuming that the MVA passes a subsequent over-training check; the 
MVA is now ready to apply to unseen data (i.e. the physics analysis).  Note that to apply an MVA to 
data used to determine HPs will result in a biased outcome and should be avoided.

\subsection{Cross Validation}
\label{sec:generalisation:cv}

The hold-out method suffers from the fact that half of the available training examples are used in order to determine the
optimal HPs.  This in turn can lead to a large variance in performance of the ML algorithm.  One can improve
upon hold-out by increasing the sample size used for HP training at the expense of the sample for validation.
This introduces the issue of how to split the data into training and test samples as the optimal HPs obtained
for a given training will depend on the data sampled from the available training examples as those examples reserved
for validation are not used in HP determination.  The $k$-fold CV algorithm divides
the data sample into $k$ distinct samples; and trains $k$ sets of HPs using a training sample that reserves the $k^{th}$ 
sub-sample (or fold) for testing, the remaining $k-1$ folds being used for training~\cite{Stone1974,Geisser1975}.  
This algorithm trains all permutations
of leaving one fold out~\footnote{This is distinct from the bootstrap approach that would permit event oversampling.}. 
The fraction of training examples used for each permutation is $(k-1)/k$.  This 
results in a reduced variance for $k>2$ compared with the hold-out approach as the $k=2$ limit corresponds to that method.
The the extreme limit
where $p$ (1) events are (is) reserved for testing is the leave $p$-out (one-out) CV method, for example
see~\cite{Arlot2010} and references therein.  To avoid
biases the training examples are independent of the test sample used to check HP generalisation prior
to applying an MVA to unseen data.

Having trained $k$ sets of HPs the temptation is to select the best performing set according
to some figure of merit (FOM), e.g. the receiver operating characteristic, ROC, integral at some pre-determined working point.
This is a biased choice for the same reason that selecting the worst case scenario is also biased.  It 
may lead to significant variation in MVA performance from one unseen data sample to the next.  The aggregate performance
of the ensemble of trainings corresponds to a more robust selection of HPs than the extremes, and the spread in performance
obtained for the different trainings contains information about algorithm variance.  Using the aggregate ensures that
the error rate corresponds to the average error rate of the $k$ folds; i.e. $E=\sum_{i=1}^k E_i / k$.
A suitable number of folds $k$ used for a particular problem needs to be determined on a case by case basis.
We have introduced $k$-fold CV to TMVA.

\subsection{Variance, Bias and testing for generalisation}
\label{sec:generalisation:other}

\begin{figure}[!ht]
\begin{center}
\includegraphics[width=0.45\columnwidth]{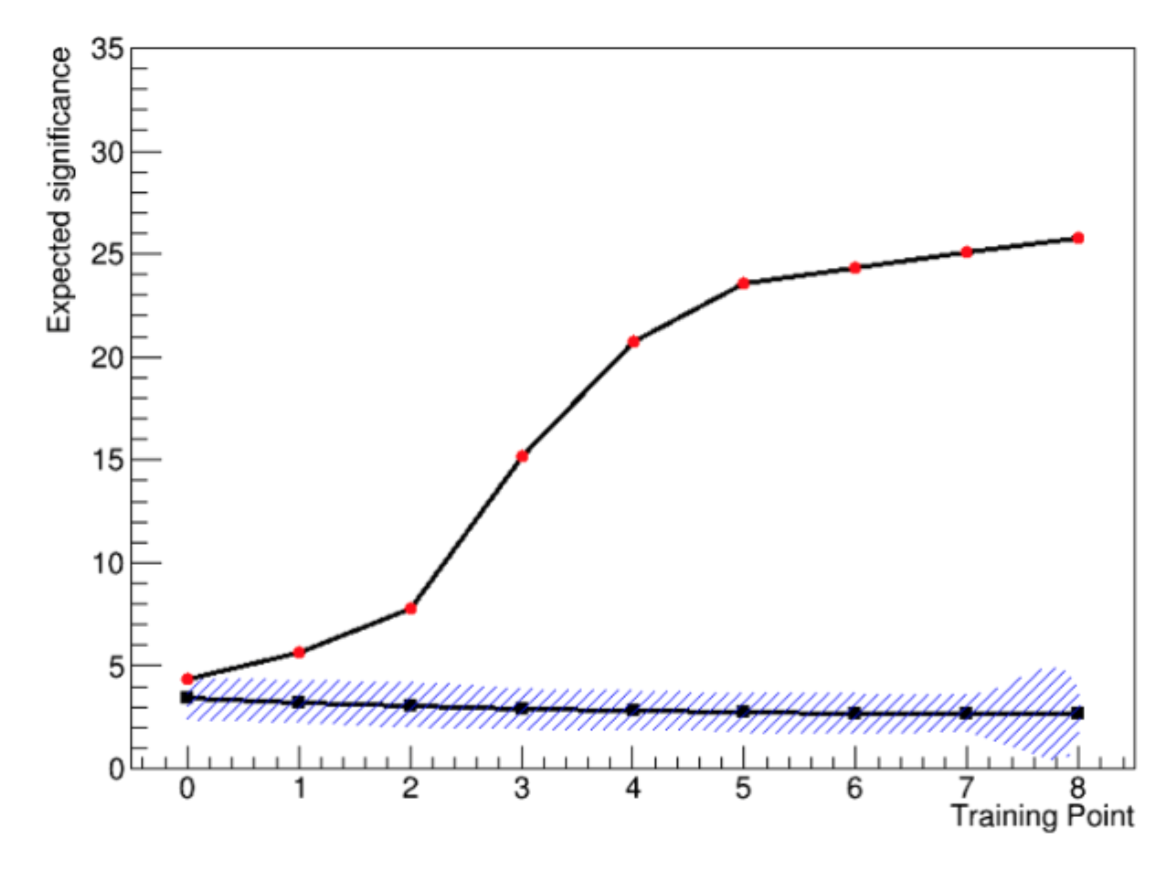}\hspace{2pc}
\begin{minipage}[b]{0.95\columnwidth}\caption{\label{fig:jackknife}Illustrating overtraining with increasing tree depth for the example in the text.}
\end{minipage}
\end{center}
\end{figure}

There are two relevant FOMs to optimise an algorithm on in particle physics, one is the
precision of a measurement and the other is significance 
of a limit obtained from a search. However often these are not used as the metric when selecting MVAs 
as a considerable effort may be required
to optimise an algorithm using one of these FOMs. Here we illustrate the situation encountered when
optimising an MVA using significance as the appropriate FOM. We apply a stratified jackknife approach 
with 50 strata and construct samples of 2000 training and 2000 validation examples obtained using 
the MadGraph Monte Carlo simulation for a new particle search.  
Training and test samples are statistically independent and each training sample is used to 
train a boosted decision tree (BDT) to distinguish between one signal and one background class of events.
To force overfitting we train BDTs with a depth between 2 and 10 (labeled as samples 0 through 8, respectively).
Figure~\ref{fig:jackknife} shows black (red) points representing the average significance obtained for the training 
(test) samples.  The band illustrated for the black points is an estimate of the standard deviation 
applied to the test sample.  The difference between red and black points indicates the level of 
HP overtraining as tree depth increases.

If the end result of a measurement is not used to construct the FOM to check for generalisation then we can consider performing a hypothesis test on the MVA output; either confusion matrices (for classification) 
or sets of histograms (for regression).  
The null and alternate hypotheses, $H_0$ and $H_1$, respectively are $H_0$: the MVA is 
generalised and $H_1$: the MVA is not generalised for all classes of event. 
For a classification problem with $2$ species of
event the fraction of (in)correctly classified species of events is indicated in the $2\times 2$ confusion matrix that 
describe the binomial classification of event type.  The parent distribution is estimated using
the average of the train and test confusion matrices.  It straight forward to construct a hypothesis
test that can be used to assert if $H_0$ is generalised at a given confidence level $1-\alpha$.  
For more than 2 species of event this approach can be extended to the corresponding multinomial problem.  For a regression 
analysis, instead of 
a confusion matrix we have a pair of histograms (one for test and one for train samples) for each species of event.  We can construct an estimator of the underlying distributions 
to perform a hypothesis test in analogy with a classification problem, where now we use a multinomial estimator for 
the fraction of events expected in each bin.

\subsection{Example: $H\to \tau^+\tau^-$}
\label{sec:example}

\begin{figure}[!ht]
\begin{center}
\includegraphics[width=0.9\columnwidth]{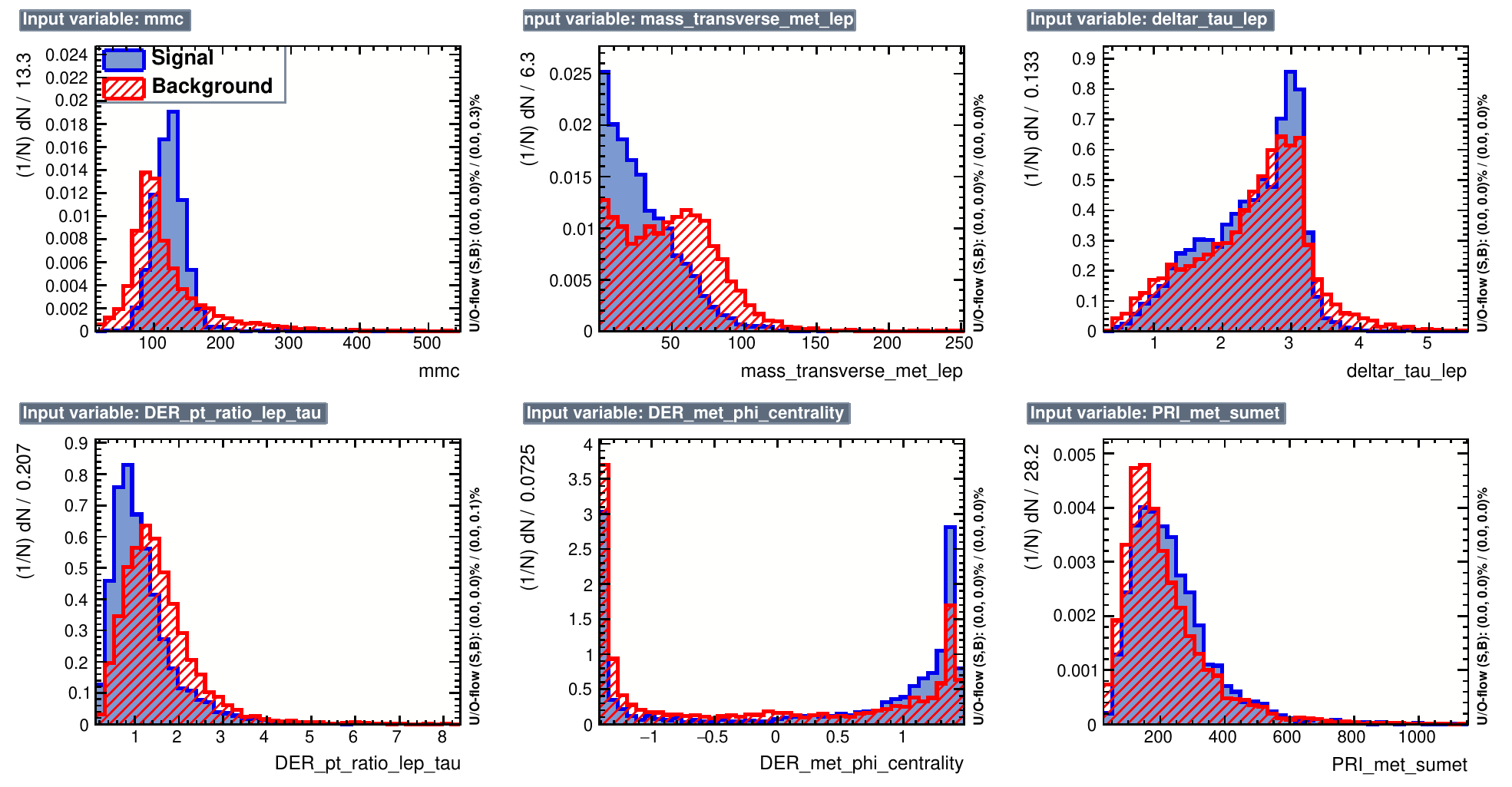}\hspace{2pc}
\begin{minipage}[b]{0.95\columnwidth}\caption{\label{fig:example:features}Distributions of input features for signal and background
for the $H\to \tau^+\tau^-$.}
\end{minipage}
\end{center}
\end{figure}

\begin{figure}[!ht]
\begin{center}
\includegraphics[width=0.45\columnwidth]{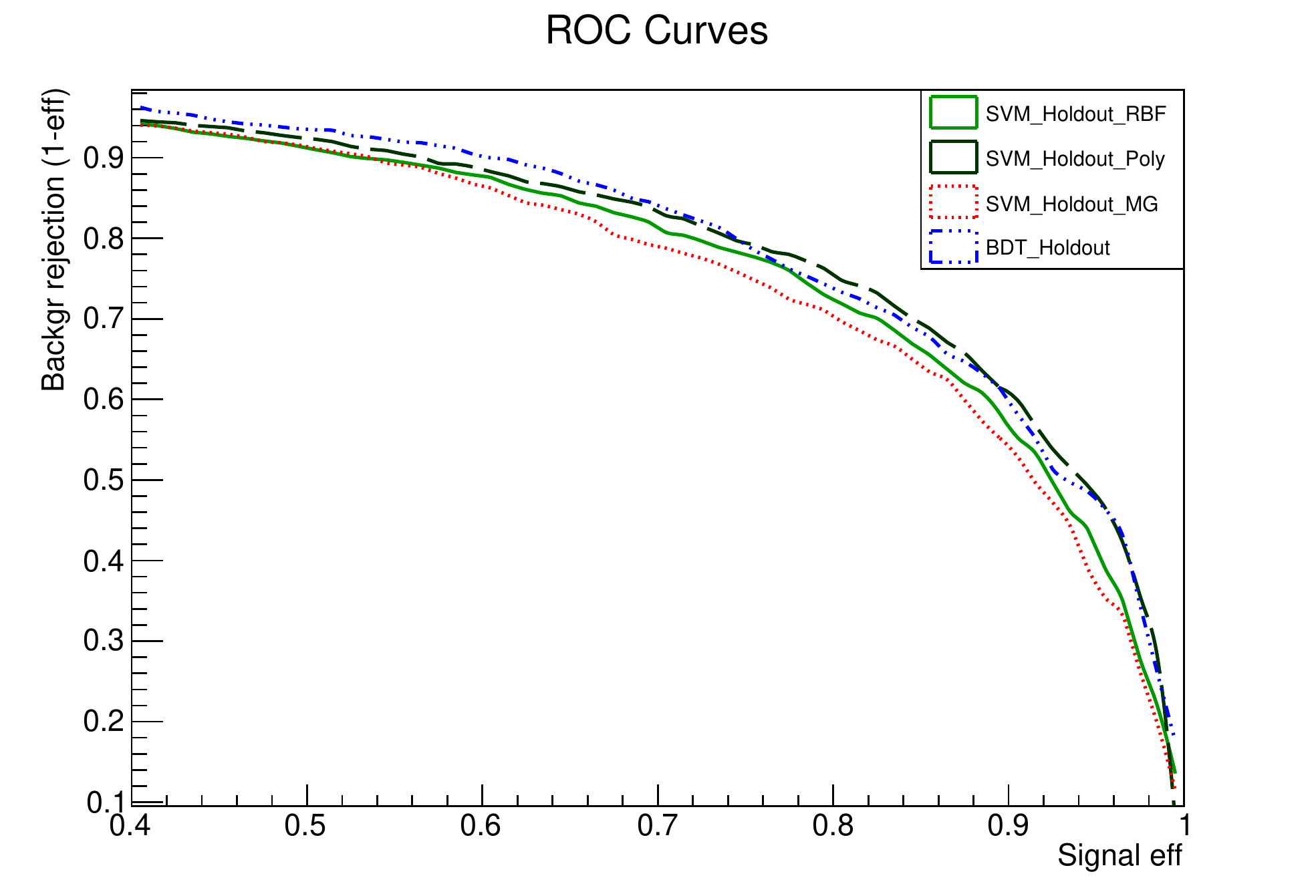}\hspace{2pc}
\includegraphics[width=0.45\columnwidth]{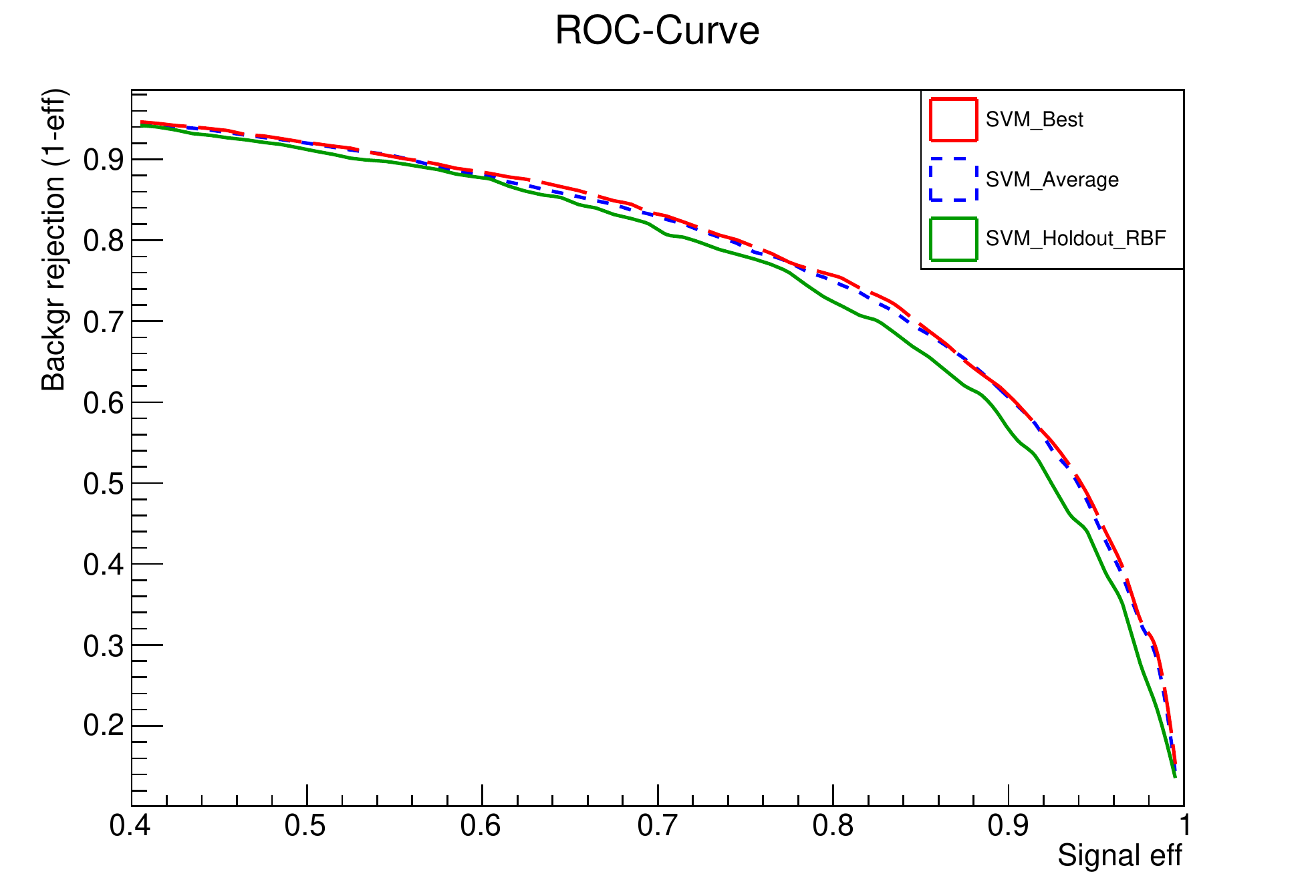}\hspace{2pc}
\begin{minipage}[b]{0.95\columnwidth}\caption{\label{fig:example:roccomparison}ROC curves for SVMs and a BDT 
as discussed in the text using (left) hold-out and (right) with 5-fold CV. The best and average SVMs determined 
in the CV process are shown to illustrate the variation between these two.}
\end{minipage}
\end{center}
\end{figure}

To illustrate the ML issues discussed above we now turn to a concrete HEP example, the search for $H\to \tau^+\tau^-$ at the LHC.
This decay has one $\tau$ decaying leptonically and the other decaying hadronically, and the final
state is not fully reconstructed due to the undetected neutrinos.
This is not a fully optimised study, but rather an illustration of potential using a limited set 
of input features from the Kaggle data set.  We use the following input features: the estimated Higgs mass using
the missing mass calculator MMC, transverse mass of the missing energy and lepton $m_T(E_T^{miss}, \ell)$, 
the $R$ separation between the hadronically and leptonically reconstructed $\tau$'s $\Delta R(\tau, \ell)$,
ratio of transverse momenta for the leptonic and hadronic $\tau$'s $p_T^\ell / p_T^\tau$,
$\phi$ Centrality, and
missing transverse energy $E_T^{miss}$.
Distributions of input features for the signal and background are shown in Figure~\ref{fig:example:features}.
This highlights that the data are overlapping and the signal features are always found in regions of
space that are also occupied by background in $X$.
These high level derived features are input to SVMs with the RBF, MG, and Poly KFs and to a BDT for performance
comparison.  All trainings use the hold-out approach and the 
corresponding ROC curves are shown in Figure~\ref{fig:example:roccomparison}.
One can see that the SVMs are able to give similar performance to the BDTs in this example; a result consistent with 
the conclusions of previous studies with SVM in HEP.  We also note that there is some variation 
in performance depending on choice of KF for this problem with the MG KF performing the worst.

We proceed to explore application of $k$-fold CV to this
problem.  The number of folds to use needs to be determined anecdotal for a given problem.  Here we illustrate the 
effect of CV on ROC integral in comparison with the hold-out technique for a 5-fold CV (see Fig.~\ref{fig:example:roccomparison}).
The best and average ROC curves are found to be similar; and while it is tempting to take the training 
corresponding to the best ROC integral to derive the nominal HP set, that is not guaranteed to provide
a typical performance for the algorithm when presented with unseen data.  The average solution corresponds
to a more robust HP choice for application of the algorithm to unseen data.  As mentioned above the 
CV method is not specific to SVMs and can be applied to other MVAs such as NNs and BDTs.  TMVA has been 
extended to include a CV tool based on this work.

Having promoted generalisation using CV one still has to address the question of HP fine tuning as these
methods do not guarantee that the optimised HP set determined correspond to a generalised result.  Likewise
when using more than one MVA, the ROC curve is not sufficient to determine which MVA leads to the best
result.  For that one has to apply the MVA to an analysis and follow through to the end result, or some 
proxy thereof; for example some FOM relating to significance such as the approximate mean significance
used in Ref.~\cite{kaggle}.

\section{Summary}
\label{sec:summary}

We have discussed the use of SVMs in the context of HEP.  Here we focus on
the example of searching for the $H\to \tau^+\tau^-$ mode at the LHC in the context of functionality available in 
TMVA.  We also discuss the issue of generalisation, highlighting some pitfalls in current practice as well as
issues to be considered when determining HPs for a given ML algorithm.  While CV is an improvement over
the hold-out method, this does not guarantee generalisation. Thus one should establish if an MVA is generalised having
previously attempted to improve the robustness of HP determination via CV or some other method.
The issues of how to illustrate overtraining by using jackknife event selection and by hypothesis testing have been 
discussed.

\end{document}